\def \as           {\hbox{$^{\prime\prime}$}}
\def \cmsq         {\hbox{cm$^{-2}$}}
\def \etal         {{\it et~al.} }
\def \Ha           {\hbox{H$\alpha$}}
\def \Hb           {\hbox{H$\beta$}}
\def \kms          {\rm{\hbox{km s$^{-1}$}}}
\def \lam          {$\lambda$}
\def \Lya          {\hbox{Ly$\alpha$}}
\def \Lyb          {\hbox{Ly$\beta$}}
\def \mum          {\hbox{$\mu$m}}
\def \pcc          {\hbox{cm$^{-3}$}}
\documentstyle[11pt,paspconf]{article}

\begin{document}

\title{Preliminary Analysis of STIS-HST Spectra of \\ 
Compact Ejecta from Eta Carinae}
\author{Fred Hamann}
\affil{Center for Astrophysics \& Space Science, University of 
California -- San Diego, La Jolla, CA 92093-0424}
\author{K. Davidson and K. Ishibashi}
\affil{Department of Astronomy, University of Minnesota, 116 Church St., 
SE, Minneapolis, MN 55455}
\author{T. Gull}
\affil{Goddard Space Flight Center, Code 680, Greenbelt, MD 20771}

\begin{abstract}
We describe some preliminary results from spatially-resolved 
spectroscopy of the compact ejection knots B and D in the 
$\eta$ Carinae wind. We emphasize various line diagnostics 
of the abundances, kinematics and physical conditions. 
The data are from new STIS-HST observations 
described by Gull \etal elsewhere in this volume. 
\end{abstract}

\keywords{spectroscopy, emission lines, Weigelt knots}

\section{Introduction}

Spectroscopic studies of the central star(s?) and complex ejecta 
of $\eta$ Car have been severely limited by the low spatial 
resolution of ground-based observations. Recent 
studies using the {\it Hubble Space Telescope} (HST) have thus 
led to an explosion of information. For example, we now 
know that the many strong and narrow emission lines that dominate 
$\eta$ Car's spectrum come from several brights knots of nebulosity 
within $\sim$0.3$^{\prime\prime}$ of the central 
star (Davidson \etal 1995, 1997). These
knots are almost certainly slow ejecta in a dense equatorial wind 
that bisects the much larger high-velocity lobes (the Homonculus, 
see also Weigelt \etal 1995, Falcke \etal 1997). 
The most prominent emission 
lines are due to singly-ionized metals, notably Fe$^+$, with 
typical velocity widths of 30--50~\kms\ (Damineli \etal 1998, 
Hamann \etal 1994).
\medskip

We obtained spatially-resolved ($<$0.1\as ) long-slit 
spectra of the star, the brightest knots and the extended 
Homonculus using the {\it Space Telescope Imaging 
Spectrograph} (STIS) -- please see T. Gull's 
description of the data elsewhere in this volume. 
The combination of high spectral resolution (20--30~\kms ) and 
wide wavelength coverage ($\sim$1650~\AA\ to $\sim$1.0~\mum ) 
allow us to employ for the first time crucial 
line diagnostics of the abundances, kinematics and 
physical conditions in spatially distinct regions. 
Below we describe some preliminary results derived 
for the bright knots B and D, with a few notes on 
applications to the direct stellar spectrum. A more 
complete analysis will appear in future papers.

\section{Reddening}

Estimates of the reddening due to dust are essential for interpretation 
of the emission line ratios and the overall spectral energy distribution. 
The simplest approach is to examine 
flux ratios of forbidden lines that arise from the same upper 
energy level of a given ion. As long as the transitions are optically 
thin, the intrinsic line ratios are then simply equal to 
$A_1\lambda_2/A_2\lambda_1$, where $\lambda_1$ and $A_1$
are respectively the wavelength and spontaneous decay rate for 
line 1, etc. Comparisons to measured ratios then provide estimates 
of the reddening (Osterbrock 1989). Among the line pairs that 
appear not to be blended with other features, we found that 
[FeII]~\lam 3175/\lam 5551, [FeII]~\lam 3533/\lam 6355 and 
[NiII]~\lam 4326/\lam 7256 yield consistent results for both knots 
B and D -- corresponding to $A_v\approx 2$ for a standard interstellar 
reddening curve (Osterbrock 1989). It is not clear that the 
reddening towards $\eta$ Car is ``standard,'' but we will use this 
result to make first-order corrections to other diagnostic line ratios 
below. 
\medskip

In the next section, we will show that the gas densities are 
probably above the critical densities of the various low-lying 
metastable levels of Fe$^+$ and Ni$^+$ that produce the 
forbidden lines. The levels 
should therefore be in collisional equilibrium, with populations 
determined by the (uncertain) gas temperature. In a forthcoming 
paper, will use this result to simultaneously employ many [FeII] 
and [NiII] lines and derive both the reddening and temperature. 
\medskip

The direct stellar spectrum in our STIS-HST data set shows some broad 
forbidden lines that apparently form in the inner stellar wind 
and not the extended ejecta. Since 
the star and knots have similar observed brightness, the extinction 
toward the star must be much larger than toward the knots. We would 
like to measure their extinctions and reddenings separately. 
Unfortunately, the only useful pair of broad forbidden lines that appears 
to be free of blends, [NII]~\lam 3063/\lam 5755, yields an unphysical 
result and so must be corrupted by blends after all.

\section{Density}

The STIS-HST spectra of knots B and D provide many forbidden 
line diagnostics of the density. 
All of them indicate densities at or near their critical limits 
for collisional deexcitation. For example, the ratios 
[SII]~\lam6716/\lam 6731 and [SII]~\lam 4069/\lam 6731 imply 
densities of $n_e\ga 10^4$ and $\ga$10$^6$~\pcc , respectively, 
based on theoretical results in Osterbrock (1989) and 
Hamann (1994). The highest densities come from 
lines of [FeII] and [NiII], with reference to calculations by 
Bautista \& Pradhan (1996). For example, the measured ratio 
[FeII]~\lam 7155/\lam 8617 implies $n_e\ga 10^7$~\cmsq , 
while [NiII]~\lam 7412/\lam 7387 and [NiII]~\lam 3439/\lam 3993 
indicate electron densities $n_e\ga 10^8$~\cmsq . There is, 
perhaps, a range of densities within the knots such that each 
of these lines forms near its critical density. 
\medskip

Not surprisingly, the measured 
ratio of broad [FeII]~\lam 7155/\lam 8617 lines in the direct 
stellar spectrum also indicates 
$n_e\ga 10^7$~\cmsq\ in the stellar wind. 

\section{Temperature}

The usual nebular diagnostics of temperature (Osterbrock 1989) 
are either too weak (for example [OIII]) or too sensitive to the 
density in this high-density environment (eg. [NII]). 

\section{Ionization}

The STIS-HST spectra were obtained near the time of the 
periodic 5.5~yr ``event,'' which is known to correspond to a 
low ionization state in the nebular emission-lines 
(Damineli 1996 and this volume). 
Thus, as expected, the doubly ionized lines such as [OIII], 
[ArIII], and [SIII] are very weak or absent. Also weak or 
absent are narrow recombination lines of HI, HeI and HeII. 
The weakness of these lines, plus the great strength of 
singly ionized lines like FeII and other discussed above, 
suggests that knots are just partially ionized -- that is, there 
is a significant amount of H$^o$ relative to H$^+$. Substantial 
partially ionized zones do not occur in normal low-density 
nebular environments that are photoionized by early type B 
stars. We will return to the significance of this point 
in \S7 below. 
Here we note that the preponderance of H$^o$ is consistent 
with the detection of narrow absorption features in the HI Balmer 
lines (see Gull \etal this volume, also below). The location of 
the absorber with respect to the knots is unknown, but its 
small velocity shift and low velocity dispersion suggest an 
association with the dense equatorial ejecta. 

\section{HI Balmer Line Absorption}

Balmer line absorption in the extended ejecta 
is interesting because it requires 
significant populations in the $n=2$ level of HI, 
10.2~eV above the ground state. If the 
local velocities are thermal and the gas temperature is 
$\sim$10,000~K, it is easy to show that optical depth 
$\tau\sim 1$ in \Hb\ requires a column density of 
$N(n=2) \ga 10^{13}$~\cmsq\ in the $n=2$ level. 
If the local velocities are larger 
due to turbulence, the column density needed for $\tau\sim 1$ 
is also larger. 
\medskip

Significant populations in $n=2$ might occur by collisions from
$n=1$ in an environment where \Lya\ is very optically thick. 
The $n=2$ level can be ``thermalized'' if the 
following condition is met,
\begin{equation}
{{n_eq}\over{A\beta}}\approx{{n_e\tau_o}\over{n_{cr}}} \ga 1
\end{equation}
where $q$ is the downward collision rate coefficient, $\tau_o$ 
is the line center optical depth in \Lya , $\beta\approx 1/\tau_o$ 
is the escape probability of \Lya\ photons, and 
$n_{cr}\approx A/q\approx 3\times 10^{17}$~\pcc\ is the critical 
density for collisional deexcitation of the $n=2$ level. 
If the product $n_e\tau_o$ is too small to satisfy this relation, 
collisions will be too infrequent to build up a significant 
$n=2$ population. For densities $n_e\la 10^9$~\pcc , Equation 1 
implies that the 
required optical depth is $\tau_o\ga 3\times10^8$ and the 
total HI column density must be 
$N(HI)\approx N(n=1)\ga 5\times 10^{21}$~\cmsq\ (again assuming 
thermal line widths). 
\medskip

The situation is actually more complicated because 
recombination into, and photoionization out of, $n=2$ 
will also effect the level population. 
Furthermore, the $2s$ level of HI is metastable -- depopulated 
by 2-photon decay but not by \Lya\ line radiation. 
The transition probability for 2-photon decay from $2s$ is just 
$\sim$8~s$^{-1}$ compared to $\sim$$5\times 10^8$~s$^{-1}$ 
for \Lya\ out of $2p$. Thus substantial $2s$ populations 
might occur without large optical depths in \Lya . 
On the other hand, collisional mixing among the $l$ states at 
high densities will work to decrease the $2s$ population. 
We will present a more thorough study of this problem 
in a forthcoming paper. 

\section{FeII, Resonant Line Pumping and Partially-Ionized Gas} 

Several metal ions are known to be photo-excited in $\eta$ Car 
by the absorption of HI Lyman line radiation. 
This resonant photoexcitation occurs via 
accidental wavelength coincidences. 
One of the most important cases involves Fe$^+$, where 
\Lya\ photons are absorbed and electrons are ``pumped'' from 
lower metastable levels into highly excited states 
(see also contributions by Johansson, 
Zethson and Davidson in this volume).  
The subsequent cascades produce a unique and sometimes 
dramatic spectral signatures. This process might actually 
dominate the overall production of 
FeII flux from $\eta$ Car and other astrophysical sources 
(Penston 1987). Measurements of primary FeII cascade lines 
show clearly that substantial \Lya\ pumping occurs 
in both the star and knots of $\eta$ Car (also Johansson \& Hamann 
1993, Hamann \etal 1994). One of us (FH) has begun a collaboration 
with G. Ferland, K. Verner and D. Verner to numerically 
simulate the FeII emission from various 
environments. Resonant line pumping is important 
only in special circumstances, and we hope to use the pumped 
FeII lines as diagnostics of the local conditions. 
\medskip

We can already draw several conclusions without detailed 
simulations. First, the metastable Fe$^+$ 
levels must be significantly populated 
and, therefore, the gas densities are probably above the 
critical densities of those levels, ie. $n_e\ga 10^6$~\pcc . 
This result is consistent with our density estimates above. 
Second, the local \Lya\ line width must be 
large because the transitions feeding some of the clearly 
pumped Fe$^+$ levels do not have good wavelength coincidences 
with \Lya . In particular, the upper level of FeII~\lam 2508 
is fed by a transition shifted $\sim$640~\kms\ from 
the \Lya\ central wavelength. This level is clearly 
pumped by \Lya , so we conclude that the \Lya\ line 
is at least $2\times 640 = 1280$~\kms\ wide. Similary, 
the fluorescent line FeII~\lam 9123 requires a \Lya\ line 
width of $2\times 670 = 1340$~\kms . Since the 
region is optically thick to \Lya\ radiation, the 
\Lya\ photons must be produced locally and the line 
width (in this otherwise low-velocity region) 
must be caused by the large optical depths. A simple 
scaling relation between the width and optical depth 
in \Lya\ (Elitzur \& Ferland 1986) suggests that 
$\tau_o\ga 2\times 10^8$ is needed to achieve the 
widths noted above (if the local Doppler velocities are roughly 
thermal and $T\approx 10,000$~K). This optical depth 
corresponds to a column density of 
$N(HI)\ga 4\times 10^{21}$~\cmsq , which is surprisingly 
similar to our estimate from the Balmer line absorption (\S6). 
This similarity is surely a coincidence, but it strengthens 
the case for large amounts of neutral hydrogen 
(\S5). Given that the knots have diameters 
$<$$4\times 10^{15}$~cm (based on angular diameters 
$<$0.1\as\ and a distance of 2300~pc to $\eta$ Car), 
we conclude that the space density in HI is 
$>$$10^6$~\pcc .
\medskip

Another interesting conclusion follows from the extensive 
work on FeII emission from quasars and active galactic 
nuclei (AGNs, eg. Kwan \& Krolick 1981, Wills \etal 1985, 
Verner \etal 1998). The FeII lines do form in ionized (HII) 
regions, but rather in partially-ionized zones behind 
the nominal HII--HI recombination front. Such zones are relatively 
small (thin) around normal stellar HII regions because they are 
very optically thick to the ionizing Lyman continuum radiation. 
But AGNs can have extensive partially-ionized zones 
because 1) penetrating X-rays from the non-thermal 
continuum source heat the gas and maintain significant 
ionization levels, and 2) high gas densities can
maintain substantial populations in the $n=2$ level of hydrogen 
(see also \S6) and thus allow photoionization by Balmer continuum 
radiation. The latter situation is also known to occur in the 
dense envelopes around luminous 
young stellar objects (Hamann \& Persson 1989 and refs. 
therein). 
\medskip

The strong FeII emission, its pumping by \Lya , and the 
Balmer line absorption (\S6) all indicate that there are 
extensive partially-ionized zones associated with the 
knots and inner ejecta of $\eta$ Car. Further evidence 
for such a region comes from measured \Lyb -pumped 
lines of OI and probably MgII in the $\eta$ Car knots 
(see also Hamann \etal 1994 and refs. therein), which requires 
optical depths in \Ha\ of at least 1000 (to keep 
\Lyb\ photons from ``leaking'' out via \Ha , 
Grandi 1980). We are planning 
photoionization simulations, with J. Hillier and 
the collaborators mentioned above, that will use the 
overall emission-line spectra of the knots to constrain both 
the local physical conditions and the spectral energy 
distribution (SED) of the central source. 
This indirect study 
of the unobservable SED could be valuable for testing 
models of the stellar wind (Hillier this volume) and 
of the single versus binary nature of the central 
object (Damineli, Davidson this volume). In particular, the 
proposed companion star should be much hotter than 
the luminous primary, dominating the 
overall SED at short wavelengths. We will try to 
determine if such a hot component is needed to 
understand the nebular line spectrum. 

\section{Abundances}

The many collisionally-excited forbidden and 
semi-forbidden (intercombination) emission 
lines from the $\eta$ Car knots 
provide numerous opportunities for abundance estimates. 
The theoretical flux ratio for any two collisionally-excited lines 
emitted from the same volume by idealized two-level atoms is, 
\begin{equation}
{{F_1}\over{F_2}} \ = \ Q \ {{n_{l1}}\over{n_{l2}}} \
{{\Omega_1\, \lambda_2\, g_{l2}}\over{\Omega_2\, \lambda_1\, g_{l1}}} 
\ e^{-{{\Delta E_{12}}\over{kT_e}}} 
\end{equation}
where $Q$ is defined by, 
\begin{equation}
Q \ \equiv \ {{1+{{n_e}\over{n_{cr2}\, \beta_2}}}\over
{1+{{n_e}\over{n_{cr1}\, \beta_1}}}}
\end{equation}
For each line 1 and 2, $F$ is the flux, 
$\lambda$ is the wavelength, $\Omega$ is the collision strength, 
$n_l$ and $g_l$ are the number density and statistical weight of 
the lower energy state, $\beta$ is the line escape probability 
($0\leq\beta\leq 1$) and $n_{cr}$ is 
the critical density. $T_e$ is the electron temperature and 
$\Delta E_{12}\equiv E_{1}-E_{2}$ is the energy 
difference between the two upper states.
\medskip

The factor $Q$ in Equation 2 corrects for possible photon 
trapping and collisional deexcitation. If the densities are low 
such that $n_e\ll n_{cr}\beta$ for both lines, then collisional 
deexcitation is not important and $Q\approx 1$. If, on the other 
hand, the densities are high or the line photons are 
significantly trapped such that $n_e\gg n_{cr}\beta$, then 
collisional deexcitation {\it is} important and 
$Q\approx n_{cr1}\beta_1/n_{cr2}\beta_2$. If the line 
photons escape freely ($\beta_1\approx \beta_2\approx 1$) 
in the high density, the correction factor 
limit is simply $Q\approx n_{cr1}/n_{cr2}$. 
The collision strengths ultimately cancel out of Equation 
2 in this limit because the levels are 
populated according to Boltzman statistics. In that case 
we have,
\begin{equation}
{{F_1}\over{F_2}} \ = 
{{A_1\, \lambda_2\, n_{1}}\over{A_2\, \lambda_1\, n_{2}}} \ 
{{g_{u1}}\over{G_{1}}} \ {{G_{2}}\over{g_{u2}}} \
e^{-{{\Delta E_{12}}\over{kT_e}}} 
\end{equation}
where $g_{u1}$ and $g_{u2}$ are the statistical weights of the 
upper states, and $G_1$ and $G_2$ are the partition 
functions and $n_1$ and $n_2$ are the space densities 
of the ions 1 and 2. This expression 
can be applied to excited-state lines and multi-level atoms 
without correction (as long as the densities are high and 
$\beta_1\approx \beta_2\approx 1$). 
\medskip

We can derive abundances from Equations 2 or 4 by noting that, 
\begin{equation}
{{n_{l1}}\over{n_{l2}}} \ \approx \ {{n_1}\over{n_2}} \ = \ 
{{f(X_1^i)}\over{f(X_2^j)}} \, \left({{X_1}\over{X_2}}\right) 
\end{equation}
where 
$f(X_1^i)$ is the fraction of element $X_1$ in ion stage $X_1^i$, 
etc., and $X_1/X_2$ is the abundance ratio by number. 
We must choose line pairs that 1) are emitted from the same or 
nearly the same region, 2) require small ionization corrections, 
and 3) have similar excitation energies so that the 
temperature-sensitive exponential factors are small. 
One strategy is to consider summed combinations of 
lines of the same element to average over these 
uncertainties. 
\medskip

Abundance studies of $\eta$ Car will help quantify the 
enrichment of the interstellar medium, test models of 
the stellar evolution and nucleosynthesis, and probe 
the the existence of dust 
within various ejecta via depletion signatures. For example, 
we are interested in the relative CNO 
abundances to gauge the amount of CNO processing of the gas. 
(The nuclear reaction rates are such that, in equilibrium, CNO 
burning of H into He converts most of the C and O into N.) 
Several UV intercombination lines are particularly valuable 
for this purpose, eg. CIII]~\lam 1909, SiIII]~\lam 1892, 
NIII]~\lam 1749 and OIII]~\lam 1664. Unfortunately, these 
and other lines from doubly-ionized species were weak 
in our on-event data (\S5). However, 
they are present in our spectra of the knots obtained in 1996 
using the HST {\it Faint Object Spectrograph} 
(FOS, Davidson \etal 1997). We find that 
N/C is roughly 25--140 times solar and N/Si is 12--63 times 
solar, depending on the density. N/O is at least 60 times 
solar for any density. From the new STIS-HST data, 
we similarly estimate Fe/O of 90--180 
times solar based on [FeII]~\lam 8617/[OI]~\lam 6300 and 
[FeII]~\lam 7155/[OI]~\lam 6300. 
These results imply that the knot gas has been extensively 
CNO processed, consistent with previous 
findings for the outer lobes (Davidson \etal 1986, and  
Dufour \etal this volume). 
We will examine the abundances in more detail in our forthcoming 
paper. 

\acknowledgments

We are grateful to Roberta Humphreys for organizing 
this stimulating and enjoyable workshop, and we thank the 
staff of the {\it Space Telescope Science Institute} for 
their kind assistance with the observations. We also acknowledge 
financial support from the HST Guest Observer program. 
FH was further supported by NASA grant NAG 5-3234.

\end{document}